# Streetlight Effect in Post-Publication Peer Review: Are Open Access Publications More Scrutinized?

Abdelghani Maddi[1], Emmanuel Monneau[2], Catherine Guaspare-Cartron [3], Floriana Gargiulo[4], Michel Dubois[5]

[1]abdelghani.maddi@cnrs.fr
ORCID : 0000-0001-9268-8022
GEMASS – CNRS – Sorbonne Université, 59/61 rue Pouchet 75017 Paris, France.
[2] manu_monneau@hotmail.com
GEMASS – CNRS – Sorbonne Université, 59/61 rue Pouchet 75017 Paris, France.
[3]catherine.guaspare@cnrs.fr
GEMASS – CNRS – Sorbonne Université, 59/61 rue Pouchet 75017 Paris, France.
[4]floriana.gargiulo@cnrs.fr
GEMASS – CNRS – Sorbonne Université, 59/61 rue Pouchet 75017 Paris, France.
[5]michel.dubois@cnrs.fr
ORCID : 0000-0001-6872-9525
GEMASS – CNRS – Sorbonne Université, 59/61 rue Pouchet 75017 Paris, France.

## Abstract

The *Streetlight Effect* represents an observation bias that occurs when individuals search for something only where it is easiest to look. Despite the significant development of Post-Publication Peer Review (PPPR) in recent years, facilitated in part by platforms such as PubPeer, existing literature has not examined whether PPPR is affected by this type of bias. In other words, if the PPPR mainly concerns publications to which researchers have direct access (eg to analyze image duplications, etc.). In this study, we compare the Open Access (OA) structures of publishers and journals among 51,882 publications commented on PubPeer to those indexed in OpenAlex database (#156,700,177). Our findings indicate that OA journals are 33% more prevalent in PubPeer than in the global total (52% for the most commented journals). This result can be attributed to disciplinary bias in PubPeer, with overrepresentation of medical and biological research (which exhibits higher levels of openness). However, after normalization, the results reveal that PPPR does not exhibit a Streetlight Effect, as OA publications, within the same discipline, are on average 16% less prevalent in PubPeer than in the global total. These results suggest that the process of scientific self-correction operates independently of publication access status.

## Acknowledgments

Data available from The Center for Scientific Integrity, the parent nonprofit organization of Retraction Watch, subject to a standard data use agreement. The authors would like to thank the Pubpeer Foundation for authorizing the collection and use of data associated with the operation of their platform.

## Competing interests

The authors have no relevant financial or non-financial interests to disclose.

## Funding information

This work was supported by a grant overseen by the French National Research Agency (ANR). Grant number: ANR-20-CE26-0008. Website: https://anr.fr/Projet-ANR-20-CE26-0008.

## 1. Introduction

Over the past decade, the evolution of Post-Publication Peer Review (PPPR), facilitated by the emergence of platforms like PubPeer (https://pubpeer.com/static/about), has led to a "deconfinement" of the process of self-correction in science (Dubois et Guaspare 2019). Historically, when applied to manuscripts already published, the scientific self-correction primarily operated through letters to editors, expressions of concern, and similar channels. The advent of platforms like PubPeer has significantly democratized and broadened the scope of PPPR. Now, anyone has potentially access to the online post-publication review process and may become a contributor, anonymous or not, providing valuable feedback and critical assessments. PPPR allows researchers and the scientific community at large to engage in discussions, identifying potential flaws, and contributing to the refinement of scientific knowledge (Jingshen 2022; Ortega et Delgado-Quirós 2023; Teixeira da Silva 2018).

The strength of PPPR stems from its apparent simplicity: it replicates the principles of a journal club in an online setting (Ortega 2022). An initiator puts forth a reading by launching a discussion thread and providing comments, setting the stage for interactions between participants and sometimes between participants and authors. Such discussions often pave the way for meticulous assessments of publications, diving deep into data, imagery, and other content (Barbour et Stell 2020). Consequently, ensuring access to these publications is of paramount importance. Since PPPR requires access to the complete content of publications, it is essential for articles to be fully available. Consequently, publications in journals accessible only through subscriptions might receive comparatively less scrutiny than those in Open Access, potentially eluding the grasp of PPPR. This hypothesis aligns with the well-known "Streetlight Effect" (Freedman 2010), wherein reviewers are naturally drawn towards easily accessible publications, creating the possibility that articles published in subscription-based journals might not undergo the same level of scrutiny and oversight. As a result, certain issues or flaws in research articles within these journals might not be thoroughly addressed, impacting the ability of the process of self-correction of science to identify and rectify potential shortcomings in scholarly publications comprehensively. Consequently, the potential influence of the well-known Streetlight Effect on PPPR cannot be overlooked.

In this study, we propose to test the Streetlight Effect hypothesis in PPPR and its potential impact on the scrutiny of scholarly publications according to their open access status. Consequently, we seek to addresses the following question: Does the open access status of publications influence the level of scrutiny they received in the context of Post-Publication Peer Review (PPPR)? Surprisingly, to date, the existing literature has not delved into this specific question. Therefore, this article aims to fill this research gap by investigating the potential impact of the Streetlight Effect on PPPR and examining whether Open Access publications undergo a more rigorous review process compared to articles accessible through subscription-based journals. Through this investigation, we seek to contribute valuable insights to the understanding of how publication accessibility may affect the thoroughness and fairness the process of self-correction of science.

To do so, we conduct a comparative analysis of publishers and journals between a dataset of 51,882 reviewed publications on PubPeer and a vast collection of 156,700,177 publications indexed in OpenAlex. Additionally, we investigate the prevalence of Open Access (OA) publishers and journals in PubPeer to explore whether the accessibility status of publications might influence their likelihood of being subjected to PPPR (measured by number of comments in PubPeer).

## 2. Data and Methods
### 2.1. PPPR data

In this study, we used PubPeer's database to analyze PPPR. PubPeer is a prominent and widely-used online platform that serves as a dynamic database for Post-Publication Peer Review (PPPR). Established in 2012, PubPeer provides a space for researchers and the scientific community to engage in open discussions, critique, and evaluation of published research articles across various disciplines (Barbour et Stell 2020; Teixeira da Silva 2018). It offers an accessible and transparent forum for individuals to post anonymous or signed comments on specific publications, addressing concerns, identifying potential flaws, and offering constructive feedback. This interactive platform allows for continuous and collaborative assessment of scholarly works, contributing to the enhancement of scientific integrity and the self-correcting nature of academic research. PubPeer's database houses a diverse range of publications, enabling researchers to actively participate in the critical appraisal and evaluation of the scientific literature.

Our initial analyses have uncovered a notable spike in comments in the year 2016, as depicted in Figure 1. These comments were observed to be generated en masse through the use of dedicated software (for further details, refer to (Baker 2016)).

In order to ensure the integrity and impartiality of our analysis, we have excluded the publications that were affected by these automated comments, amounting to a total of 47,633 publications (publications with only a single comment from commentator No. 3845).

**Figure 1: Number of publications per year before and after excluding automatically generated comments**

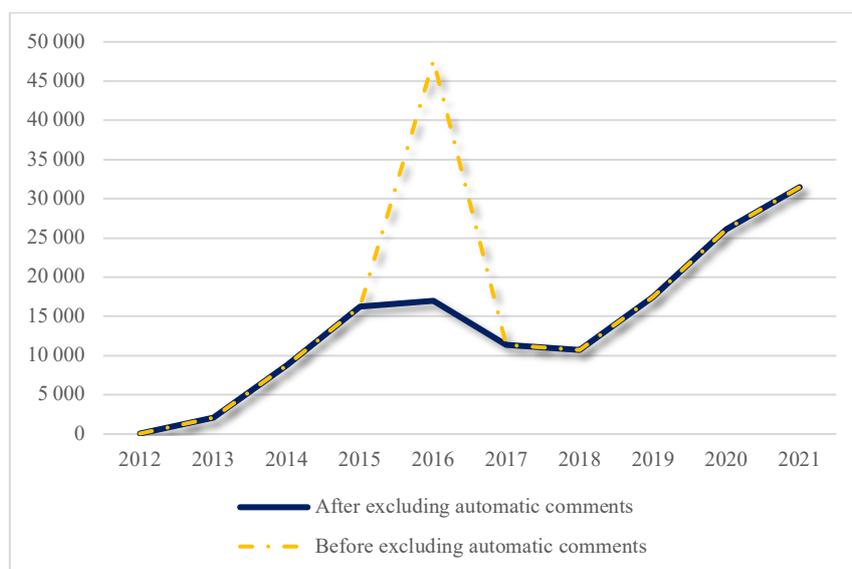

Consequently, our final sample of publications subject to comments in PubPeer consists of 51,882 articles, enabling us to conduct a robust and unbiased examination of the peer-reviewed literature in the context of Post-Publication Peer Review.

### 2.2. OpenAlex data

OpenAlex is a comprehensive and extensive database that offers broad coverage of academic publications across various disciplines (Priem, Piwowar, et Orr 2022; Singh Chawla 2022). With a vast collection of indexed materials, OpenAlex includes research articles, conference papers, preprints, and other scholarly outputs from a diverse range of sources. This database is designed to encompass a wide spectrum of publication types and supports, facilitating access to a rich array of academic content for researchers and scholars. By curating and aggregating data from numerous publishers and repositories, OpenAlex provides a valuable resource for researchers seeking a comprehensive view of scholarly literature and enables in-depth investigations across different fields of study.

In this study, we utilized the "openalexR" package (https://docs.ropensci.org/openalexR/) to extract crucial information regarding journals and publishers from the OpenAlex database. The extracted dataset comprises essential details, including the journal's name, the name of its publisher, the Open Access (OA) status of the journal, and its inclusion in the Directory of Open Access Journals (DOAJ). Additionally, the dataset includes the number of publications and citations associated with each journal.

In total, the extracted dataset encompasses information from 253,222 diverse supports of publication, including directories, open archives, and more. Out of these, 186,347 entries correspond to journals, which constituted the focus of our analysis in this paper. The dataset's breadth and depth provided us with a comprehensive overview of the publishing landscape, enabling us to conduct a thorough investigation into the relationship between Open Access publications and the Post-Publication Peer Review process.

### 2.3. Normalized Open Access Index

The Normalized Open Access Index (NOAI) is a metric utilized to quantify the extent of Open Access adoption within a specific entity, such as an institution, region, or country, following the conventional method of normalization in bibliometrics (for further details, refer to (Maddi 2020)). This metric employs a normalized scale, with a value of 1 serving as the neutral reference point, representing the global average. This normalization technique facilitates equitable comparisons of Open Access integration across diverse disciplines, accounting for disciplinary disparities in the prevalence and availability of Open Access publishing. We calculated the NOAI using data at the level of Journal Categories from the Web of Science (WOS) database, which provide more reliable data on disciplinary classification.

Thus, before proceeding with the normalization process, we first conducted a matching procedure between the publications commented on PubPeer and the publications available in the Web of Science (WoS) database, utilizing Digital Object Identifiers (DOIs) as identifiers. As a result of this matching process, a total of 46,775 (90%) out of the initial 51,882 commented publications were successfully matched with corresponding entries in the WoS database.

The normalization process involves two steps. Firstly, we calculated the share of Open Access (OA) publications per subject category within PubPeer and then divided it to the corresponding share at the global level (in the WoS database). This step ensures that the OA adoption in PubPeer is assessed relative to the broader publishing landscape. In the second step, to obtain an overall normalized indicator of OA for PubPeer, we computed a weighted average by considering the number of publications per discipline. This approach accounts for variations in the volume of OA publications across different fields and results in the Normalized Open Access Index (NOAI).

### 3. Results

In this section we present the results of our analysis. To commence, we outline a comprehensive depiction of the market structure within the scholarly publishing marked, as indexed in the OpenAlex database. This analysis involves delineating the market shares of the top publishers and constructing a Lorenz curve that illustrates the skewed distribution of publishing activity. Notably, this analysis is conducted at a holistic level rather than segmented by specific disciplines. Subsequently, we turn our focus to the publishers that have garnered the highest levels of commentary on PubPeer. Moreover, we delve into the most commented-upon journals and provide an examination of their Open Access status, thereby offering a holistic understanding of their place on the post-publication peer review process. Lastly, we present the outcomes derived from our analysis of the Normalized Open Access Index (NOAI), a crucial measure that helps rectify the disciplinary bias present in PubPeer, which exhibits an overrepresentation of life sciences publications (which tend to be more Open Access-oriented than other disciplines). Through the exploration of these diverse dimensions, our objective is to provide a comprehensive analysis aimed at determining whether, in the context of Post-Publication Peer Review (PPPR), reviewers exhibit a propensity to concentrate their focus on Open Access publications and journals or not. This investigation seeks to uncover potential patterns or biases in the post-publication review process, shedding light on whether the accessibility status of publications influences their likelihood of undergoing scrutiny within the PPPR framework. By examining these intricate dynamics, we endeavor to contribute valuable insights into the interplay between Open Access adoption, peer review practices, and the broader landscape of scholarly communication.

### 3.1. Landscape of scholarly publishing marked

Table 1 offers a static yet illuminating snapshot of the current landscape of scholarly publishing, providing a glimpse into the prevailing market dynamics. Among the total of 10,097 publishers analyzed, the top 26 publishers emerge as pivotal players, collectively accounting for a substantial 60% share of the publications indexed in OpenAlex. This concentration underscores the considerable influence wielded by a relatively small number of publishers in shaping the dissemination of academic knowledge.

Within this landscape, Elsevier BV takes center stage with a dominant 17.4% share, while Springer Nature follows closely at 9.2%. Wiley commands a significant 8.5% share, and Taylor & Francis is represented at 3.3%. These findings underscore the established presence of these industry giants within the scholarly publishing domain. Notably, the emergence of newcomers like the Multidisciplinary Digital Publishing Institute (MDPI) adds an intriguing dimension to this tableau. While MDPI has gained traction and achieved an impressive 11th position, it is important to acknowledge that its practices have been subject to critique by a substantial majority of open access advocates. MDPI's publication of numerous special issues and other aggressive strategies have raised concerns and skepticism within the scholarly community, drawing attention to ongoing debates surrounding responsible open access practices.

**Table 1: Top 26 publishers (out of 10 097) in OpenAlex**

| N° | Publishers | Works count | Market share | Cumulative share |
|---|---|---|---|---|
| 1 | Elsevier BV | 19 742 706 | 17,4% | 17,4% |
| 2 | Springer Nature | 10 418 723 | 9,2% | 26,7% |
| 3 | Wiley | 9 674 047 | 8,5% | 35,2% |
| 4 | Taylor & Francis | 3 705 323 | 3,3% | 38,5% |
| 5 | Oxford University Press | 3 633 513 | 3,2% | 41,7% |
| 6 | SAGE Publishing | 2 801 120 | 2,5% | 44,2% |
| 7 | Lippincott Williams & Wilkins | 2 172 203 | 1,9% | 46,1% |
| 8 | Cambridge University Press | 2 013 390 | 1,8% | 47,9% |
| 9 | American Chemical Society | 1 907 889 | 1,7% | 49,6% |
| 10 | Institute of Electrical and Electronics Engineers (IEEE) | 1 204 259 | 1,1% | 50,6% |
| 11 | Multidisciplinary Digital Publishing Institute (MDPI) | 1 135 521 | 1,0% | 51,6% |
| 12 | BMJ | 1 000 571 | 0,9% | 52,5% |
| 13 | American Institute of Physics | 882 739 | 0,8% | 53,3% |
| 14 | IOP Publishing | 795 996 | 0,7% | 54,0% |
| 15 | Royal Society of Chemistry | 780 349 | 0,7% | 54,7% |
| 16 | De Gruyter | 742 071 | 0,7% | 55,3% |
| 17 | American Physical Society | 709 382 | 0,6% | 56,0% |
| 18 | American Medical Association | 703 057 | 0,6% | 56,6% |
| 19 | Thieme Medical Publishers (Germany) | 627 903 | 0,6% | 57,1% |
| 20 | University of Chicago Press | 521 330 | 0,5% | 57,6% |
| 21 | Emerald Publishing Limited | 493 780 | 0,4% | 58,0% |
| 22 | Chinese Medical Association | 478 360 | 0,4% | 58,5% |
| 23 | Frontiers Media | 460 097 | 0,4% | 58,9% |
| 24 | Routledge | 445 724 | 0,4% | 59,3% |
| 25 | Karger Publishers | 444 252 | 0,4% | 59,6% |
| 26 | American Association for the Advancement of Science | 416 783 | 0,4% | 60,0% |

As Table 1 captures the current publishing landscape, it serves as a foundation for deeper explorations into the intricate dynamics that shape scholarly communication. This analysis contributes to our understanding of the evolving interplay between established publishing giants, emerging contenders, and the ongoing discussions regarding the ethical and strategic dimensions of open access publishing (Nicholas et al. 2023; Yamada et Teixeira da Silva 2022).

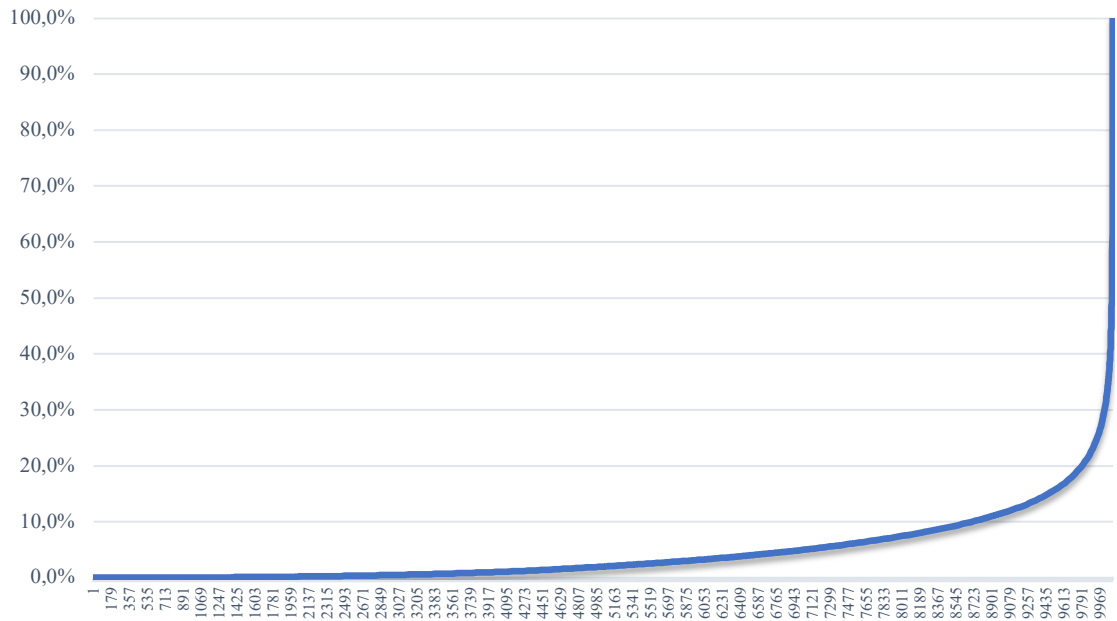

**Figure 2: Lorenz curve of the publishing market, according to OpenAlex database**
Gini index = 0,91

The Lorenz curve (Figure 2), reveals a highly asymmetric distribution within the scholarly publishing market. The curve, which showcases the cumulative share of publications against the cumulative share of publishers, underscores a pronounced concentration of influence among a select few. With an accompanying Gini coefficient of 0.91, this curve conveys a staggering level of inequality in the distribution of publishing activities.

The steep upward slope of the Lorenz curve signifies that a disproportionately large portion of publications is concentrated among a handful of prominent publishers. This pattern is indicative of a highly centralized landscape, where a limited number of publishing entities wield substantial control over the dissemination of academic knowledge (Larivière, Haustein, et Mongeon 2015; Roy et Yami 2006). The wide divergence from the hypothetical line of perfect equality demonstrates the substantial variance between publishers, accentuating the market's lopsided nature.

As the scholarly community grapples with questions of access, equity, and influence, the striking asymmetry illustrated by the Lorenz curve, coupled with the elevated Gini coefficient both serve as a powerful landmark for fostering discussions about the impact of market concentration and its implications for open access, diversity of voices, and the democratization of knowledge dissemination (Bernius et al. 2009; Ponte, Mierzejewska, et Klein 2017).

### 3.2. The most commented publishers in PubPeer

**Table 2: Most commented publishers in PubPeer***

| Publishers | # Commented publications in Pubpeer | Works count (OpenAlex) | Share in PubPeer (1) | Share in total (OpenAlex) (2) | Relative Share RS=1/2 | Share of commented publication in the publisher's total |
|---|---|---|---|---|---|---|
| Verduci Editore (European Review for Medical and Pharmacological Sciences) | 691 | 10 724 | 1,33% | 0,01% | 140,53 | 6,4% |
| Cognizant Communication Corporation | 140 | 3 154 | 0,27% | 0,00% | 96,81 | 4,4% |
| Impact Journals LLC | 967 | 34 684 | 1,86% | 0,03% | 60,81 | 2,8% |
| Cell Press | 1 104 | 77 134 | 2,13% | 0,07% | 31,22 | 1,4% |
| Spandidos Publishing | 768 | 62 653 | 1,48% | 0,06% | 26,73 | 1,2% |
| e-Century Publishing Corporation | 191 | 16 684 | 0,37% | 0,01% | 24,97 | 1,1% |
| Landes Bioscience | 304 | 26 832 | 0,59% | 0,02% | 24,71 | 1,1% |
| Ivyspring International Publisher | 130 | 14 489 | 0,25% | 0,01% | 19,57 | 0,9% |
| American Society for Biochemistry and Molecular Biology | 1 975 | 223 770 | 3,81% | 0,20% | 19,25 | 0,9% |
| American Association for Cancer Research | 1 216 | 142 756 | 2,34% | 0,13% | 18,58 | 0,9% |
| eLife Sciences Publications Ltd | 120 | 14 657 | 0,23% | 0,01% | 17,86 | 0,8% |
| American Society for Clinical Investigation | 265 | 39 450 | 0,51% | 0,03% | 14,65 | 0,7% |
| Dove Medical Press | 540 | 83 201 | 1,04% | 0,07% | 14,16 | 0,6% |
| IOS Press | 798 | 140 064 | 1,54% | 0,12% | 12,43 | 0,6% |
| Public Library of Science | 1 810 | 327 689 | 3,49% | 0,29% | 12,05 | 0,6% |
| National Academy of Sciences | 807 | 158 510 | 1,56% | 0,14% | 11,10 | 0,5% |
| Society for Neuroscience | 226 | 44 720 | 0,44% | 0,04% | 11,02 | 0,5% |
| Springer Nature | 8 777 | 10 418 723 | 16,92% | 9,21% | 1,84 | 0,4% |
| American Psychological Association | 1 273 | 345 487 | 2,45% | 0,31% | 8,04 | 0,4% |
| American Phytopathological Society | 165 | 45 791 | 0,32% | 0,04% | 7,86 | 0,4% |
| Rockefeller University Press | 221 | 62 788 | 0,43% | 0,06% | 7,68 | 0,4% |
| Cold Spring Harbor Laboratory Press | 141 | 42 123 | 0,27% | 0,04% | 7,30 | 0,3% |
| The Company of Biologists | 203 | 71 358 | 0,39% | 0,06% | 6,20 | 0,3% |
| Association for Research in Vision and Ophthalmology | 192 | 71 035 | 0,37% | 0,06% | 5,89 | 0,3% |
| American Association of Immunologists | 248 | 103 880 | 0,48% | 0,09% | 5,21 | 0,2% |
| American Society for Microbiology | 770 | 326 526 | 1,48% | 0,29% | 5,14 | 0,2% |
| American Diabetes Association | 143 | 61 183 | 0,28% | 0,05% | 5,10 | 0,2% |
| BioMed Central | 722 | 344 587 | 1,39% | 0,30% | 4,57 | 0,2% |
| Portland Press | 289 | 148 919 | 0,56% | 0,13% | 4,23 | 0,2% |
| American Society of Hematology | 288 | 181 671 | 0,56% | 0,16% | 3,46 | 0,2% |
| American Association for the Advancement of Science | 589 | 416 783 | 1,14% | 0,37% | 3,08 | 0,1% |
| **Hindawi Publishing Corporation** | 457 | 350 049 | 0,88% | 0,31% | 2,85 | 0,1% |
| Federation of American Societies for Experimental Biology | 158 | 124 979 | 0,30% | 0,11% | 2,76 | 0,1% |
| Frontiers Media | 551 | 460 097 | 1,06% | 0,41% | 2,61 | 0,1% |
| Informa | 389 | 397 235 | 0,75% | 0,35% | 2,14 | 0,1% |

\* Publishers with +117 publications (according to Fisher's discretization), and RS>1.5.

Table 2 provides a comprehensive overview of publishers whose publications have garnered the highest number of commentaries within the PubPeer platform. To delineate this selection, we employed a discretization approach utilizing the Fisher algorithm. This methodology facilitated the identification of a threshold, leading us to focus on publishers whose works have exceeded a critical threshold of 117 commented publications. This careful approach ensures a concentrated analysis of the most discussed publishers within PubPeer.

Moreover, our analysis goes beyond sheer volume by introducing the concept of the Relative Share (RS), a dual ratio that measures the presence of publishers in PubPeer in relation to their representation within the broader OpenAlex dataset. The RS ratio, with a neutral value of 1, allows for a nuanced comparison. For instance, an RS value of 3 indicates that a publisher's presence in PubPeer is threefold higher than its representation within the total OpenAlex corpus. The publishers featured in Table 2 meet both criteria: a publication count surpassing 117 and an RS exceeding 1.5.

By employing this multidimensional approach, Table 2 presents a snapshot of the publishers that have elicited extensive commentary within the PubPeer platform. Notably, a predominant majority of the most discussed publishers in PubPeer hail from the field of life sciences, with a particular focus on medicine and biosciences. Many of these publishers have embraced the open access model, exemplified by Verduci Editore, where the "European Review for Medical and Pharmacological Sciences" has garnered commentary on PubPeer. Similarly, e-Century Publishing Corporation and the controversial Hindawi Publishing Corporation and Frontiers Media also stand out, exhibiting respective presence ratios (RS) of 2.8 and 2.6, indicating their significantly higher prominence within PubPeer compared to OpenAlex database.

Among the major players in the top 20, Springer Nature captures attention with a substantial 16.92% of the publications commented on in PubPeer. This translates to an RS of 1.84, indicating its 84% higher presence in PubPeer relative to the total. While an observable trend emerges, showcasing a stronger representation of fully open access publishers among the most discussed in PubPeer, it is essential to exercise caution in attributing this pattern solely to their open access status. Instead, the orientation toward life sciences disciplines emerges as a critical driving factor, a phenomenon extensively discussed in the existing literature. The marked focus of Post-Publication Peer Review (PPPR) on life sciences disciplines significantly influences the prominence of certain publishers, underscoring the broader disciplinary underpinnings that shape engagement and discourse within the PPPR framework.

### 3.3. The most commented journals in PubPeer

In this subsection, our focus shifts towards a comprehensive examination of scholarly journals within the context of our study. This exploration encompasses two primary phases. Firstly, we delve into the realm of open access publishing, analyzing the prevalence of fully open access journals within the OpenAlex database, their representation within PubPeer, and their prominence among the most discussed journals on the PubPeer platform. Secondly, we embark on an analysis of the distribution of journals with the highest volumes of commented publications on PubPeer, subsequently examining their distribution based on their open access status. Furthermore, our investigation extends to journals with the most significant Relative Share (RS), an indicator that gauges their prominence within PubPeer relative to their representation in the broader OpenAlex dataset.

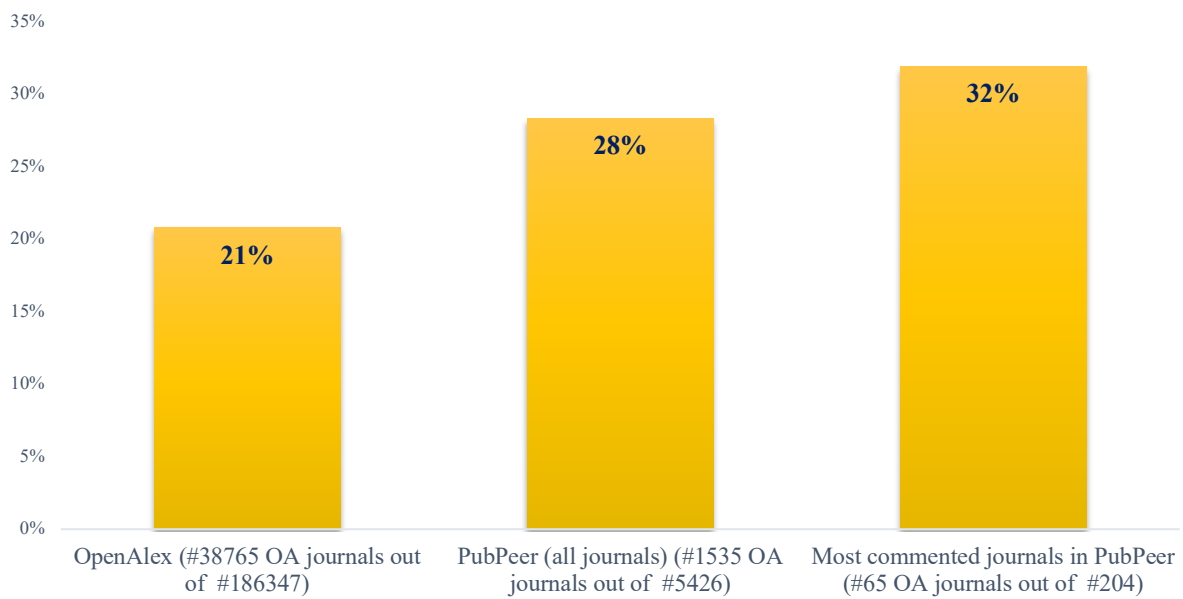

Figure 3: Share of Open Access journals, by dataset

At first glance, these results appear to suggest a distinct focus of Post-Publication Peer Review (PPPR) on open access journals. However, it is imperative to exercise caution in drawing definitive conclusions, as this apparent trend could be influenced by various factors. Indeed, these findings lay the groundwork for deeper exploration, and as we delve into the normalization process (Section 4), a more nuanced and comprehensive understanding will emerge. Through this critical analysis, we aim to disentangle the intricate dynamics that shape the PPPR process, shedding light on the interplay between open access status, peer review engagement, and the broader landscape of scholarly communication.

Table 3: Most commented journals in PubPeer

| Journals | Is OA | # Commented publications in PubPeer | Share in PubPeer (1) | Share in OpenAlex (2) | Relative Share RS=(1)/(2) |
|---|---|---|---|---|---|
| *Journal of Biological Chemistry* | True | 1951 | 3,760% | 0,131% | 28,61 |
| *PLOS ONE* | True | 1556 | 2,999% | 0,180% | 16,68 |
| *Oncotarget* | True | 842 | 1,623% | 0,017% | 93,67 |
| *Proceedings of the National Academy of Sciences of the United States of America* | False | 807 | 1,555% | 0,101% | 15,38 |
| *European Review for Medical and Pharmacological Sciences* | False | 691 | 1,332% | 0,007% | 194,61 |
| *Cancer Research* | False | 649 | 1,251% | 0,055% | 22,95 |
| *Journal of Intelligent and Fuzzy Systems* | False | 623 | 1,201% | 0,007% | 177,28 |
| *Oncogene* | False | 588 | 1,133% | 0,014% | 81,60 |
| *Molecular and Cellular Biology* | False | 429 | 0,827% | 0,020% | 40,94 |
| *Cell* | False | 416 | 0,802% | 0,016% | 49,84 |
| *Nature Communications* | True | 407 | 0,784% | 0,034% | 23,25 |
| *Journal of Cellular Biochemistry* | False | 377 | 0,727% | 0,010% | 70,76 |
| *Biomedicine & Pharmacotherapy* | True | 333 | 0,642% | 0,010% | 61,38 |
| *Cell Death and Disease* | True | 329 | 0,634% | 0,006% | 112,05 |
| *Journal of Cellular Physiology* | False | 305 | 0,588% | 0,012% | 48,47 |

Table 3 offers a focused glimpse into the landscape of scholarly journals, spotlighting the 15 journals that have garnered the highest volume of commented publications within the PubPeer platform. An intriguing observation surfaces as the data unfolds: the three most commented-upon journals all belong to the open access category. These prominent journals—Journal of Biological Chemistry, PLOS ONE, and Oncotarget. Impressively, within the expansive array of 5,426 commented journals on PubPeer, this trio alone commands attention, collectively constituting more than 8% of the total commented publications.

Furthermore, an additional observation comes to light from this table: a substantial majority of the featured journals align with the life sciences domain. This consistent trend serves to reinforce the earlier observation we gleaned from Table 2, albeit on a more granular level, pertaining to publishers.

**Table 4: Most commented journals in PubPeer (with the highest RS)**

| Journals | Is OA | # Commented publications in PubPeer (n >=100) | Share in PubPeer (1) | Share in OpenAlex (2) | Relative Share RS=(1)/(2) |
|---|---|---|---|---|---|
| Artificial Cells Nanomedicine and Biotechnology | True | 199 | 0,384% | 0,001% | 284,18 |
| Oncology Research | False | 140 | 0,270% | 0,001% | 227,70 |
| European Journal of Psychology of Education | False | 108 | 0,208% | 0,001% | 211,13 |
| European Review for Medical and Pharmacological Sciences | False | 691 | 1,332% | 0,007% | 194,61 |
| Journal of Intelligent and Fuzzy Systems | False | 623 | 1,201% | 0,007% | 177,28 |
| Journal of Experimental & Clinical Cancer Research | True | 148 | 0,285% | 0,002% | 124,72 |
| Cancer Cell | False | 136 | 0,262% | 0,002% | 114,48 |
| Cell Death and Disease | True | 329 | 0,634% | 0,006% | 112,05 |
| Molecular Cancer | True | 111 | 0,214% | 0,002% | 111,60 |
| Molecular Plant-microbe Interactions | True | 157 | 0,303% | 0,003% | 102,73 |
| Cellular Physiology and Biochemistry | True | 216 | 0,416% | 0,004% | 102,26 |
| Cell Metabolism | False | 118 | 0,227% | 0,002% | 98,89 |
| OncoTargets and Therapy | True | 197 | 0,380% | 0,004% | 93,89 |
| Oncotarget | True | 842 | 1,623% | 0,017% | 93,67 |
| Bioscience Reports | True | 171 | 0,330% | 0,004% | 89,20 |

Table 4 presents a selection of the most commented journals, with the highest relative share (RS) scores. While occupying a distinctly modest stature within the OpenAlex database, these journals display a demonstrably increased presence relative to their weight on the PubPeer platform.

An important observation emanates from this table: a significant majority of these widely reviewed journals align with the open access paradigm, accounting for 9 of the top 15 entries (or 60%). This trend aligns with the patterns we explored in Tables 2 and 3, further accentuating the prominence of open access journals in the body of commented publications on PubPeer. In addition, an overarching thematic thread connects these results to the focus of these journals, with a clearly dominant representation of medical and bioscience publications within PubPeer.

### 3.4. The Normalized Open Access Index (NOAI)

In the previous sections, we examined the composition of publishers and journals within the PubPeer platform and the comprehensive OpenAlex database, highlighting the prevalence of open access entities. This section goes further in terms of granularity by examining the composition of commented publications on PubPeer, according to their open access status. This analytical phase represents a deliberate effort to mitigate the disciplinary bias evident in the earlier analyses, which might suggest an inherent tendency for open access journals to attract more commentary within PubPeer. As elucidated in the "Methods" section, our application of the Normalized Open Access Index (NOAI) serves as a potent tool in unraveling the intricate interplay between open access practices and post-publication peer review.

To compute the Normalized Open Access Index (NOAI), our methodology unfolds through a two-step process. In the initial phase, we ascertain the share of open access (OA) publications within PubPeer ($OA_{ij}/x_{ij}$), categorized according to the WoS Subject Categories. This share is subsequently normalized by relating it to the proportion of each WoS Subject Category within the comprehensive Web of Science (WoS) database ($OA_{wj}/X_{wj}$). This initial normalization procedure brings to light the distribution of open access ($OA_{S_{ij}}$) engagement across distinct disciplinary domains within the PubPeer platform.

$$OA_{S_{ij}} = \frac{OA_{ij}/x_{ij}}{OA_{wj}/X_{wj}}$$

Subsequently, in the second phase, obtain a comprehensive indicator of openness within PubPeer by calculating a weighted average. This entails a meticulous computation that involves the weighted aggregation of the dual ratios derived in the preceding step. These dual ratios offer a revealing perspective, showcasing the relative prevalence of OA publications within PubPeer against their representation within the global WoS framework. By conducting this intricate aggregation, we derive a robust measure—the NOAI—that encapsulates the multifaceted openness dynamics operative within the post-publication peer review ecosystem. Through this methodical approach, we aspire to unravel the intricate relationships between open access practices, peer review engagement, and disciplinary orientations within the scholarly landscape.

$$NOAI_i = \frac{\sum(OA_{S_{ij}} \times x_{ij})}{x_i} = \mathbf{0.84}$$

The culmination of our analysis yields a NOAI value of 0.84. This metric conveys a nuanced revelation: open access publications, when normalized for disciplinary equivalency, exhibit a relative underrepresentation of approximately 16% within PubPeer compared to their prevalence within the global corpus of publications. This consequential insight serves to refute the Streetlight Effect hypothesis within the context of Post-Publication Peer Review (PPPR). The NOAI's discernible departure from parity elucidates that PPPR in PubPeer transcends the constraints of the Streetlight Effect, emphasizing its capacity to uphold its review integrity across diverse realms of scholarly publishing, regardless of open access status. This finding underscores the robustness and impartiality of PPPR in PubPeer in fostering rigorous scholarly scrutiny and engagement.

4. **Conclusion**

Through this article, our aim was to delve into the potential existence of the Streetlight Effect within the realm of Post-Publication Peer Review (PPPR). In other words, we sought to investigate whether reviewers display a propensity to scrutinize open access publications more extensively due to their greater accessibility. To address this inquiry, we meticulously analyzed a corpus of 51,882 commented publications within the PubPeer platform. Specifically, we juxtaposed the structural composition of publishers and journals within this subset against the broader OpenAlex and Web of Science databases.

In the absence of normalization, our initial findings revealed a discernible concentration of open access journals within PubPeer, vis-à-vis the wider publishing landscape. Additionally, our results underscored a notable prevalence of life sciences disciplines among the commented publications within PubPeer. Consequently, the pronounced prevalence of open access journals could potentially emanate from this disciplinary bias, as the open access practice exhibits a heightened prevalence within these specific domains.

The calculated Normalized Open Access Index (NOAI) serves to corroborate this hypothesis. Upon normalization, we observed that open access publications, when discipline-normalized, were relatively underrepresented by approximately 16% within PubPeer as compared to the global publication corpus. This substantiates the notion that the skew towards open access journals is largely a byproduct of the underlying disciplinary landscape, where the concentration of life sciences disciplines significantly influences the open access prevalence.

5. **Discussion**

Our investigation uncovers a nuanced narrative that illuminates the potential interplay between accessibility and post publication peer review engagement. Our findings, guided by a comprehensive analysis, contribute to the broader understanding of PPPR dynamics, revealing that the bias introduced by the Streetlight Effect may be mitigated through discipline-normalized assessments. As we continue to traverse the evolving terrain of scholarly communication, analyzing the effect of accessibility on willingness to be reviewed post-publication, this study underscores the significance of impartial and rigorous peer review processes in sustaining the integrity of the scientific discourse.

These findings carry several implications from a research policy standpoint. Firstly, the results, for the first time, demonstrate that Post-Publication Peer Review (PPPR) operates independently of publication access status, dispelling concerns of bias from this perspective. In essence, publications, once published, undergo consistent scrutiny irrespective of their accessibility, suggesting that the review process remains resilient to access constraints. The notion of "pirate libraries" like Sch-hub emerges intriguingly as a potential facilitator of reviewer access to scientific contents, prompting further inquiry into their potential role. This avenue presents a compelling direction for future exploration.

Secondly, a noteworthy outcome is the reduced concentration of open access (OA) publications within PubPeer, following normalization. This observation hints at the possibility that OA publications may be subject to fewer issues, warranting a deeper investigation. In essence, this poses the critical question of whether the openness of publications, encompassing research data as well, contributes to bolstering scientific integrity. In other words, to what extent does the opening of publications (including research data) contribute to the strengthening of scientific integrity? This question beckons for future exploration and could serve as a catalyst for shaping research policy and practices.

In conclusion, these results reverberate across the landscape of scholarly communication and research policy, challenging preconceived notions and stimulating a reconsideration of the interplay between open access, peer review dynamics, and self-correction of science. As we chart the course ahead, these insights underscore the multifaceted nature of the scholarly enterprise, prompting us to delve deeper into the mechanisms that underpin peer review, publication accessibility, and their collective contribution to the advancement of knowledge.

## 6. Avenues for Future Research

An intriguing avenue for future research could involve exploring another facet of the "Streetlight Effect" not addressed in this article. This time, the focus would be on investigating whether researchers tend to concentrate their Post-Publication Peer Review (PPPR) efforts on articles published in the most renowned or dominant journals in their field, potentially at the expense of less well-known or marginal journals. This form of research bias could potentially have implications for the quality and objectivity of PPPR, as prestigious journals often garner more attention and funding, which could influence the frequency of comments and critiques. By examining whether such a "Spotlight Effect" exists in the context of PPPR, researchers could gain a better understanding of how publication dynamics influence the processes of scientific evaluation and correction while aiming to promote a more balanced assessment of academic literature.